\documentstyle[12pt,aasms4]{article}

\received{23 September 1998}

\accepted{10 December 1998}

\lefthead{Bottke et al.}

\righthead{Asteroids Shaped by Planetary Tides}

\begin{document}

\title{1620 Geographos and 433 Eros: Shaped by Planetary Tides?}

\author{W. F. Bottke, Jr.}

\affil{Center for Radiophysics and Space Research, Cornell University, Ithaca,
NY 14853-6801}

\author{D. C. Richardson}

\affil{Department of Astronomy, University of Washington, Box 351580, Seattle,
WA 98195}

\author{P. Michel}

\affil{Osservatorio Astronomico di Torino, Strada Osservatorio 20, 10025 Pino
Torinese (TO), Italy}

\and

\author{S. G. Love}

\affil{Jet Propulsion Laboratory, California Institute of Technology, M/S
306-438, 4800 Oak Grove Drive, Pasadena, CA 91109-8099}







\begin{abstract}

Until recently, most asteroids were thought to be solid bodies whose shapes
were determined largely by collisions with other asteroids (\cite{dav89}).  It
now seems that many asteroids are little more than rubble piles, held together
by self-gravity (\cite{bur98}); this means that their shapes may be strongly
distorted by tides during close encounters with planets. Here we report on
numerical simulations of encounters between a ellipsoid-shaped rubble-pile
asteroid and the Earth. After an encounter, many of the simulated asteroids
develop the same rotation rate and distinctive shape (i.e.,  highly elongated
with a single convex side, tapered ends, and small protuberances swept back
against the rotation direction) as 1620 Geographos.  Since our numerical
studies show that these events occur with some frequency, we suggest that
Geographos may be a tidally distorted object. In addition, our work shows that
433 Eros, which will be visited by the NEAR spacecraft in 1999, is much like
Geographos, which suggests that it too may have been molded by tides in the
past.

\end{abstract}






\keywords{asteroids, dynamics, celestial mechanics}










%



\section{Introduction to 1620 Geographos}

The shapes of several Earth-crossing objects (ECOs) have now been inferred by
delay-Doppler radar techniques (\cite{ost93,ost95a,ost95b,hud94,hud95,hud97}).
They show that ECOs have irregular shapes, often resembling beat-up potatoes
or even contact binaries. It is generally believed that these shapes are
by-products of asteroid disruption events in the main belt and/or cratering
events occurring after an ECO has been ejected from its immediate precursor. 
A few of these bodies, however, have such unusual shapes and surface features
that we suspect an additional reshaping mechanism has been at work.  As we
will show, at least one ECO, 1620 Geographos, has the exterior
characteristics, orbit, and rotation rate of an object which has been
significantly manipulated by planetary tidal forces.  

1620 Geographos is an S-class asteroid with a mean diameter slightly over 3
km.  It was observed with the Goldstone 2.52-cm (8510-MHz) radar from August
28 through September 2, 1994 when the object was within 0.0333 AU of Earth
(\cite{ost95a,ost96}). A delay-Doppler image of Geographos's pole-on
silhouette (Fig.\ 1) showed it to have more exact dimensions of $5.11 \times
1.85$ km ($2.76 \times 1.0$, normalized), making it the most elongated object
yet found in the solar system (\cite{ost95a,ost96}).  In addition,
Geographos's rotation period ($P = 5.22$ h) is short enough that loose
material is scarcely bound near the ends of the body (\cite{bur75}). For
reference, Geographos would begin to shed mass for $P \lesssim 4$ h if its
bulk density was 2.0 g ${\rm cm}^{-3}$ (Harris 1996; Richardson et al. 1998).

\placefigure{fig1}

Geographos's elongated axis ratio was unusual enough that Solem and Hills
(1996) first hypothesized it may not be a consequence of collisions. Instead,
they speculated it could be a by-product of planetary tidal forces, which
kneaded the body into a new configuration during an encounter with Earth.

To test their hypothesis, they employed a numerical $N$-body code to track the
evolution of  non-rotating strengthless spherical aggregates making close slow
passes by the Earth.  Some of their test cases showed that tidal forces
stretch spherical progenitors into cigar-like figures as long or longer than
the actual dimensions of Geographos.  Since ECOs undergo close encounters with
Earth (and Venus) with some frequency (Bottke et al., 1994), Solem and Hills
(1996) postulated that other ECOs may have comparable elongations.

Though Geographos's elongation is provocative, it is, by itself, an inadequate
means of determining whether the asteroid has been modified by tidal forces. 
To really make the case that 1620 Geographos is a tidally distorted object,
several questions must be answered: 

\begin{enumerate}

\item[A.] Is Geographos's internal structure (or that of any other ECO) weak
enough to allow tidal forces to pull it apart? 

\item[B.] How likely is it that Geographos ever made a close slow encounter
with a large terrestrial planet like Earth or Venus?

\item[C.] Can tidal forces reshape an ECO into a Geographos-like silhouette
(not just an asymmetrical elongated figure) and reproduce its spin rate?

\item[D.] If so, how often do such events occur?  

\item[E.] Is Geographos a singular case, or have other ECOs undergone
comparable distortion?

\end{enumerate}

In the following sections, we will address each of these questions in turn.
Our primary tool to investigate these issues is the $N$-body code of
Richardson et al. (1998), more advanced than the code of Solem and Hills
(1996) and capable of determining the ultimate shape and rotation of our
progenitors.  By applying a reasonable set of ECO starting conditions, we will
show that Geographos-type shapes and spins are a natural consequence of tidal
disruption. The results discussed here are based on the extensive parameter
space surveys completed for Richardson et al. (1998). 

\section {Issue A: Evidence that ECOs are ``rubble piles''}

Planetary tidal forces are, in general, too weak to modify the shapes of solid
asteroids or comets unless the bodies are composed of very weak material
(\cite{jef47,opi50}). Recent evidence, however, supports the view that most
km-sized asteroids (and comets) are weak ``rubble-piles'', aggregates of
smaller fragments held together by self-gravity rather than material strength
(\cite{cha78}, \cite{lov96}).  We list a few salient points; additional
information can be found in Richardson et al. (1998). (i) Comet
Shoemaker-Levy-9 (SL9) tidally disrupted when it passed within 1.6 planetary
radii of Jupiter in 1992; numerical modelling suggests this could only have
happened if SL9 were virtually strengthless (\cite{asp96}). (ii) C-class
asteroid 253 Mathilde has such a low density (1.3 g ${\rm cm}^{-3}$;
\cite{vev98}) compared to the inferred composition of its surface material
(i.e., if carbonaceous chondrite-like, it would have a density $\sim 2$ g
${\rm cm}^{-3}$; Wasson 1985) that its interior must contain large void
spaces,  small fragments with substantial interparticle porosity, or a
combination of the two. (iii) A set of 107 near-Earth and main belt asteroids
smaller than 10 km shows no object with a rotation period shorter than 2.27 h;
this spin rate matches where rubble-pile bodies would begin to fly apart from
centrifugal forces (\cite{har96}). (iv) Most collisionally evolved bodies
larger than $\sim 1$ km are highly fractured, according to numerical
simulations of asteroid impacts (\cite{asp93,gre96,lov96}). (v) All of the
small asteroids (or asteroid-like bodies) imaged so far by spacecraft (e.g.,
253 Mathilde, 243 Ida, 951 Gaspra, and Phobos) have large craters on their
surface, implying their internal structures are so broken up that they damp
the propagation of shock waves and thereby limit the effects of an impact to a
localized region (\cite{asp98}). If this were not the case, many of these
impacts would instead cause a catastrophic disruption.

If the above lines of evidence have been properly interpreted, we can conclude
that Geographos (and other ECOs) are probably rubble piles, since ECOs are
generally thought to be collisionally evolved fragments derived from
catastrophic collisions in the main belt. Thus, we predict that Geographos is
weak enough to be susceptible to tidal distortion during a close pass with a
planet.

\section {Issue B: Probable Orbital Evolution of 1620 Geographos}

If Geographos is a tidally distorted object, it had to encounter a planet at
some time in the past. Not just any encounter will do, however. Tidal forces
drop off as the inverse cube of the distance between the bodies, such that
distant encounters far outside the planet's Roche limit cause negligible
damage to the rubble pile. High velocity trajectories past a planet leave
little time for tidal forces to modify the rubble pile's shape. Thus, we need
to estimate the probability that Geographos has made a close slow encounter
with Earth or Venus. 

 The orbits of ECOs evolve chaotically.  Many of them have orbits which allow
them to encounter multiple planets and the terrestrial planet region is
crisscrossed with secular and mean-motion resonances (\cite{fro95,mic97b}).
For these reasons, it is impossible to accurately track the orbital motion of
any ECO more than a few hundred years into the past or future. The only way,
therefore, to assess the likelihood that Geographos had a planetary encounter
in the past is to numerically integrate its orbit with that of many clones, in
the hope that broad evolution patterns can be readily characterized. To this
end, following the procedure of Michel et al.,\ (1996), we used a
Bulirsch-Stoer variable step-size integration code, optimized for dealing
accurately with close encounters, to track the evolution of 8 Geographos-like
test clones. We integrated the nominal orbit with $a=1.246$ AU, $e=0.335$,
$i=13.34^{\circ}$;  the other clones were defined by slightly changing their 
orbital parameters one at a time. All of the planets were included except
Pluto. Orbital parameters were provided by the JPL's Horizons On-line
Ephemeris System v2.60 (Giorgini et al. 1998). Each clone was followed for 4
Myr.

In general, we determined the orbital evolution of the clones to be controlled
by two mechanisms: close encounters with Earth and overlapping secular
resonances $\nu_{13}$ and $\nu_{14}$ involving the mean precession frequencies
of the nodal longitudes of Earth and Mars's orbits (Michel and Froeschl\'e
1997, Michel 1997). We found that 5 of the 8 clones (62.5\%) had their
inclinations increased by these resonances.  This trend opens the possibility
that these mechanisms could have affected Geographos's orbit in the past and
consequently that its inclination has been pumped up from a lower value.
Similarly, 6 of the 8 clones (75\%) had their orbital eccentricities increased
by the $\nu_2$ and $\nu_5$ secular resonances with Venus and Jupiter. Lower
eccentricities and inclinations in the past imply that close approaches near
Earth were even more likely to occur, and to happen at the low velocities
conducive for tidal disruption, in agreement with integrations by other groups
(\cite{fro95}). Thus, these integrations moderately increase our confidence
that Geographos has been stretched by tides in the past.

\section {Issue C. Tidal Disruption Model and Results}

\subsection {The model}

To investigate the effects of planetary tides on ECOs (cf., \cite{sol96}), we
have used a sophisticated $N$-body code to model Earth flybys of
spherical-particle aggregates  (\cite{bot97,bot98,der98}). Our goal in this
section is to determine whether Geographos-like shapes are common by-products
of tidal disruption.  Model details, analysis techniques, and general results
described in Richardson et al. (1998). For brevity, we only review the basics 
here.

The particles' motions are tracked during the encounter using a 4th-order
integration scheme that features individual particle timesteps (Aarseth 1985).
This method allows us to treat interparticle collisions rigorously, with a
coefficient of restitution included to produce energy loss (i.e.,  friction);
previous models usually assumed elastic or perfectly inelastic collisions.
Note that if energy dissipation is not included, clumps formed by
gravitational instability are noticeably less tightly bound (\cite{asp96}). 

The code is capable of modelling tidal disruption over a range of rubble pile
shapes, spin rates, spin-axis orientations, and hyperbolic trajectories.  To
verify the code was accurate enough to realistically model shape changes, we
consulted two experts in granular media, J. Jenkins of Cornell University, and
 C. Thornton of Aston University, UK.  Based on their suggestions, we checked
our code against some standard diagnostic tests in their field.  For our first
test, we numerically modeled spherical particles being dropped into a pile
along a flat surface. Our results showed that we were able to reproduce an
empirically-derived angle of repose. For a second test, we examined the pre-
and post-planetary encounter particle configurations of our rubble piles to
determine whether their shapes were artifacts of a crystalline lattice
structure (i.e., ``cannonball stacking'').  Our results showed that lattice
effects are nearly unavoidable in rubble pile interiors, especially when
same-sized spherical particles are used, but that the outer surfaces of our
rubble piles had essentially randomized particle distributions. Thus, based on
our success with these tests and the positive comments of the granular media
experts, we have some confidence that our $N$-body code yields reasonable
results. 

Our model rubble piles had dimensions of $2.8 \times 1.7 \times 1.5$ km, our
choice for a representative ECO shape (Richardson et al. 1998), and bulk
densities of 2 g cm$^{-3}$, similar to the estimated densities for Phobos and
Deimos (\cite{tho92}). Note that this value may be overly-conservative,  given
the 1.3 g cm$^{-3}$ density found for Mathilde.  Individual particles have
densities of 3.6 g cm$^{-3}$, similar to ordinary chondritic meteorites
(\cite{was85}). For most test cases, our rubble pile consisted of 247
particles, with each particle having a diameter of 255 m. Same-sized particles
were chosen for simplicity; future work will investigate more plausible
particle size-distributions. Cases deemed interesting were examined further
using rubble piles with 491 same-sized particles. In these instances, particle
densities were modified to keep the aggregate's bulk density the same as
before.  We found that the change in resolution did not significantly modify
the degree of mass shedding, the final shape, or the final spin rate of the
model asteroid, though it did make some shape features more distinctive. 

The tidal effects experienced during a rubble pile's close approach to Earth
are determined by the rubble pile's trajectory, rotation, and physical
properties.  To investigate such a large parameter space, Richardson et al.
(1998) systematically mapped their outcomes according to the asteroid's
perigee distance $q$ (between 1.01 and 5.0 Earth radii), approach speed
$v_\infty$ (between 1.0 and 32 km s$^{-1}$), rotation period $P$ (tested at
$P$ = 4, 6, 8, 10, and 12 h for prograde rotation, $P =  6$ and 12 h for
retrograde rotation, and the no-spin case $P = \infty$), spin axis orientation
(obliquity varied between $0^\circ$ and $180^\circ$ in steps of $30^\circ$),
and orientation of the asteroid's long axis at perigee (tested over many
angles between $0^\circ$ and $360^\circ$).  We discuss the outcomes,
especially those pertaining to Geographos, below.

\subsection {Tidal disruption outcomes}

Several distinct outcomes for tidal disruption were found by Richardson et al.
(1998). The most severely disrupted rubble piles were classified as ``S'', a
``SL9-type'' catastrophic disruption forming a line of clumps of roughly equal
size (a ``string of pearls'') with the largest fragment containing less than
50\% of the progenitor's original mass.  Less severe disruptions were
classified as ``B'', break-up events where 10\% to 50\% of the rubble pile was
shed into clumps (three or more particles) and single particles. Mild
disruption events were classified as ``M'', with the progenitor losing less
than 10\% of its mass. As we will show below, each outcome class is capable of
producing Geographos-like elongations and spin rates. 

\subsection{Reshaping rubble piles with planetary tidal forces}

To quantify shape changes, we measured the length of each rubble pile's axes
after encounter ($a_1 \ge a_2 \ge a_3$), calculated the axis ratios ($q_2
\equiv a_2 / a_1$ and $q_3 \equiv a_3 / a_1$), and defined a single-value
measure of the remnant's ``ellipticity'' ($\varepsilon_{{\rm rem}} \equiv 1 -
\frac{1}{2} (q_2 + q_3)$). For reference, our progenitor has 
$\varepsilon_{{\rm rem}} = 0.43$ and Geographos has a value of
$\varepsilon_{{\rm rem}} = 0.64$. 

Sampling a broad set of parameters to map tidal disruption outcomes,
Richardson et al. (1998) identified 195 S, B, or M-class events produced with
a $\varepsilon_{{\rm rem}} = 0.43$ rubble pile. Fig.\ 2 shows this set with
ellipticity plotted against the fraction of mass shed by the progenitor during
tidal disruption. 

\placefigure{fig2}

We find that, in general, S-class events tend to yield lower ellipticity
values; only 2 of the 79 outcomes are likely to have a Geographos-like
elongations ($\varepsilon_{{\rm rem}} > 0.60$). The mean value of
$\varepsilon_{{\rm rem}}$ for the S-class events is 0.22 with standard
deviation $\sigma$ = 0.14.  The near-spherical shapes produced by S-class
events are a by-product of gravitational instabilities in the fragment chain
which readily agglomerate scattered particles as they recede from the planet.

B-class events do not show a simple trend with respect to ellipticity, though
these values tend to increase as the degree of mass shedding decreases. We
find that 5 of 40 outcomes have Geographos-like $\varepsilon_{{\rm rem}}$
values. The mean value of $\varepsilon_{{\rm rem}}$ for all 40 B-class events
is 0.45 ($\sigma = 0.14$), very close to the starting ellipticity of 0.43. 

M-class events are most effective at increasing $\varepsilon_{{\rm rem}}$ and
creating Geographos-like shapes, probably because tidal torques must first
stretch and/or spin-up the rubble pile before particles or clumps can be
ejected near the ends of the body. Fig.\ 2 shows 23 of 76 M-class events with
Geographos-like $\varepsilon_{{\rm rem}}$ values.  Overall, the 76 outcomes
have a mean $\varepsilon_{{\rm rem}}= 0.54$ with $\sigma = 0.10$. Thus,
getting a Geographos-like ellipticity from a M-class disruption is less than a
1 $\sigma$ event, decent odds if such disruptions (and $\varepsilon_{{\rm
rem}} = 0.43$ progenitors) are common.

\subsection{Spin-up and down with planetary tidal forces}

Tidal disruption also changes the spin rates of rubble piles. This can be done
by applying a torque to the non-spherical mass distribution of the object,
redistributing the object's mass (and thereby altering its moment of inertia),
removing mass (and angular momentum) from the system, or some combination of
the three. Fig.\ 3 shows the spin periods of the remnant rubble piles ($P_{\rm
rem}$) for the 195 disruption cases described above. Recall that the range of
starting $P$ values was 4, 6, 8, 10, and 12 h for prograde rotation, $P =  6$
and 12 h for retrograde rotation, and the no-spin case $P = \infty$. 

The mean spin period for 79 S-class outcomes is $5.6 \pm 2.2$ hours, while the
comparable value for the 40 B-class and 76 M-class events is $5.2 \pm 1.1$ and
$4.9 \pm 1.1$ hours, respectively.  Note that these last two values are close
to the real spin period of Geographos (5.22 h).  These similar values indicate
that mass shedding only occurs when the km-sized bodies are stretched and
spun-up to rotational break-up values. The final rotation rate of the rubble
pile is then  determined by the extent of the mass loss; in general, more mass
shedding (S-class events) means a loss of more rotational angular momentum,
which in turn translates into a slower final spin rate.  Though the points of
Fig.\ 3 do show some scatter, the 195 disruption events together have a mean
$P_{\rm rem}$ value of $5.2 \pm 1.7$ hours, a good match with Geographos once
again.

\placefigure{fig3}

\subsection {Matching the shape and spin of 1620 Geographos}

Now that we have found tidal disruption outcomes which match Geographos's
ellipticity and spin rate, we can take a closer look at the resultant shapes
of the rubble piles themselves. Our goal is to find distinctive features which
match comparable features on Geographos, and which are  possibly antithetical
with a collisional origin. To make sure we can resolve these features, we have
 used a rubble pile containing nearly twice the number of components as before
(491 particles).  Fig.\ 4 shows this body going through a M-class event with
the following encounter parameters: $P = 6$ hours prograde, $q = 2.1
R_\oplus$, and $v_\infty = 8$ km s$^{-1}$. 

\placefigure{fig4}

Fig.\ 4a shows the asteroid before encounter.  The spin vector is normal to
the orbital plane and points directly out of the page. The asteroid's
equipotential surface (to which a liquid would conform) is a function of local
gravity, tidal, and centrifugal terms. At this stage, it hugs the outer
surface of the rubble-pile.

Fig.\ 4b shows the body shortly after perigee passage.  Here, the
equipotential surface becomes a more elongated ellipsoid with its longest axis
oriented towards Earth.  Differential tidal forces, greatest at perigee, and
centrifugal effects combine to set the particles into relative motion,
producing a landslide towards the ends of the body.  Particles above the new
angle of repose roll or slide downslope to fill the ``low spots'', and thereby
further modify the body's potential. As a consequence, the rubble pile is
elongated and, as the planet pulls on the body, its rotation rate altered. The
action of the Earth stretches the model asteroid and, by pulling on the
distorted mass, spins it up, increasing its total angular momentum.  Mass
ejection occurs when the total force on a particle near the asteroid's tips is
insufficient to provide the centrifugal acceleration needed to maintain
rigid-body rotation.

Fig.\ 4c shows the latter stages of the landslide. Particles near the tips are
swept backward in the equatorial plane by the asteroid's rotation. The
material left behind frequently preserves this spiral signature as cusps
pointing away from the rotation direction.  Note that these cusps are easy to
create but difficult to retain with identical spherical particles at this
resolution; we believe that real rubble piles, with rough or craggy
components, would more readily ``freeze'' in position near the ends. Particle
movement along the long axis is not uniform; shape changes, increased angular
momentum, and mass shedding cause one side of the body to become bow-like. 
This effect produces a convex surface along the long axis and a ``hump''-like
mound of material on the opposite side.

Fig.\ 4d shows the final shape of the object.  The spin ($P_{\rm rem}=5.03$ h)
and ellipticity ($\varepsilon_{{\rm rem}} = 0.65$) are virtually identical to
Geographos (Fig.\ 1).  The shapes of the two ends are, surprisingly, not
symmetric. We believe this is caused by the  starting topography, which can
play a decisive role in the effectiveness of tidal deformation. The strength
of tidal and centrifugal terms depends on each particle's position
(\cite{ham96}), such that some particles lie further above the local angle of
repose than others. Since our model asteroid, like real ECOs, is neither a
perfect ellipsoid nor a readily-adaptable viscous fluid, the new distorted
shape is influenced by the body's granular nature (i.e., friction and
component size affect the strength of the landslide).  Hence, particles leak
more readily off one end than the other, often accentuated by limited particle
movement before the rubble-pile reaches perigee. The end that sheds more mass
frequently becomes elongated, tapered, and narrow when compared to the
stubbier antipode. The overall final shape of the body is much like that of a
``porpoise'' or ``schmoo''. A comparison between Fig.\ 1 and Fig.\ 4 shows a
good match; all of Geographos's main features have been reproduced.

\section {Issue D. Production Rate of Geographos-Shaped Objects}

As described above, certain S-, B- and M-class disruptions can leave rubble
piles with highly elongated shapes and fast spin rates.  To estimate the
frequency of those particular disruption events near Earth and Venus, we use
the technique of Bottke et al. (1998) and combine a ``map'' of tidal
ellipticity results (described in Richardson et al., 1998) with probability
distributions based on ECO spins, ECO spin axis orientations, ECO close
approaches with Earth and Venus, and ECO encounter velocities with Earth and
Venus. Our results show that a typical ECO should undergo an S-, B-, or 
M-class  event once every $\sim 65$ Myr, comparable to an ECO's collision rate
with Earth and Venus (Richardson et al., 1998).  Similarly, this same body
should get an Geographos-like ellipticity ($\varepsilon_{{\rm rem}} > 0.60$)
once every $\sim 560$ Myr.  The most likely disruption candidates have low
$e$'s and $i$'s, consistent with the Geographos's probable orbital history
(i.e., Sec.\ 3). Since the dynamical lifetime of ECOs against planetary
collision, comminution, or ejection by Jupiter is thought to be on the order
of 10 Myr (\cite{gla97}), we predict that $\sim 15$\% of all ECOs undergo S-,
B- or M-class disruptions (i.e.,  10 Myr / 65 Myr), and that $\sim 2$\% of all
ECOs (i.e.,  10 Myr / 560 Myr) should have shapes (and spins) like Geographos.
 The implications of this prediction will be discussed below.

\section {Issue E. Other Geographos-Like Objects} 

\subsection{Detecting tidally-distorted ECOs}

Our estimate that 2\% of all rubble pile ECOs should have Geographos-type
shapes and spin periods is, at best, only accurate to a factor of several,
given the many unknown quantities we are modeling and the relatively unknown
shape distribution of the ECO population.  Still, the following thought
experiment is useful in providing a crude ``reality check''.  1620 Geographos
has a mean diameter of 3 km and an absolute magnitude of $H$ = 15.6 (Giorgini
et al. 1998).  Morrison (1992) estimates there are roughly 100 ECOs with
absolute magnitudes brighter than 15.0 (6 and 3 km diameters, respectively,
for the dark C's and bright S's).  Since 2\% of 100 objects is 2 objects, it
is perhaps not surprising we have not noticed more Geographos-like asteroids.

Alternatively, one could argue that, given these odds, it was fortunate to
have discovered Geographos's shape in the first place, especially when one
considers that only 35\% of the $H < 15.0$ ECOs have been been discovered, and
relatively few of them have had their shapes determined by delay-Doppler radar
(Morrison 1992). It is useful to recall, however, that the known ECO
population is biased towards objects which pass near the Earth on low
inclination orbits (Jedicke 1996), exactly the class of objects which are
favored to undergo tidal disruption. Hence, the discovery of Geographos's
shape among a limited sample of ECOs may not be a fluke.  We predict, though,
that more Geographos-like objects are lurking among the undiscovered ECO
population. 

Our investigation of Geographos led us to examine a second asteroid, 433 Eros,
which shares many of Geographos's distinguishing characteristics. We believe
Eros may also be tidally distorted, as we will discuss further below.

\subsection{Application to 433 Eros}

Our success in suggesting an explanation for Geographos has led us to consider
the next most elongated asteroid, S-class asteroid 433 Eros, the target of the
NEAR mission.  Eros has many of the same distinguishing characteristics as
Geographos (and our B- and M-class remnant rubble piles). Visual and radar
observations taken during a 0.15 AU pass near Earth in 1975 report that Eros
has a short rotation period (5.27 hours) and a highly elongated shape ($36
\times 15 \times 13$ km; $2.77 \times 1.2 \times 1.0$, normalized; ellipticity
$\varepsilon_{{\rm rem}} = 0.61$) (\cite{zel76,mcf89,mit98}).  Both values are
comparable to those recorded for Geographos and with 15\% (30 out of 195) of
our S-, B-, and M-class disruption cases.

Even more intriguing, however, is Eros's pole-on silhouette, which, after
modeling the older Goldstone radar data, looks something like a kidney bean
(Fig.\ 5) (\cite{mit98}).  One must be careful not to overinterpret this
shape, since it is based on data that has a signal to noise ratio of $\sim 70$
while the shape has been ``fit'' to a reference ellipsoid which can eliminate
discriminating features.  In fact, the concave side of the ``kidney bean''
shape may not be a single concavity, but several adjacent ones.  Still, we
believe it plausible that Eros's arched back and tapered ends are analogous to
similar features on Geographos, themselves produced by spiral deformations
associated with tidal forces.  Images from the NEAR spacecraft should readily
resolve this issue.   

\placefigure{fig5}

The NEAR spacecraft will offer several additional ways to test our hypothesis.
 Regardless of whether Eros is covered by regolith or bare rock, spectroscopic
measurements will suggest a surface composition which can be directly compared
to terrestrial rock samples. If the densities of these samples are
substantially larger than Eros' bulk density, we can infer that Eros is
probably a rubble pile. While observations of large craters would support the
rubble pile scenario, too many would weigh against the tidal disruption
scenario; global landslides caused by a relatively recent tidal disruption
event should modify or bury craters.  For this reason, we expect most tidally
distorted objects to have relatively young and spectroscopically uniform
surfaces. As we will describe below, however, the unknown dynamical history of
Eros makes any prediction problematic.  Landslides also sort debris as it goes
downhill; high resolution images near the ends of Eros may not only show
cusp-like features but a prevalence of small fragments. An estimate of the
spatial distribution of block sizes inside Eros may come from NEAR's gravity
field maps. Finally, the results of Bottke and Melosh (1996) and Richardson et
al., (1998) show that asteroids affected by tides may often have small
satellite companions which were torn from the original body.  Thus, the
presence of a small moon about Eros would be a strong indication that it had
undergone tidal fission.

A possible problem, dynamically-speaking, is that Eros is currently an Amor
asteroid on a solely Mars-crossing orbit ($a = 1.46$ AU, $e= 0.22$, $i=
10.8^\circ$). Test results show that tidal disruption events occur relatively
infrequently near Mars, since it is a weak perturber (Bottke and Melosh 1996).
 Studies of Eros's orbital evolution, however, suggest that it may have been
on a low-inclination, deeply Earth-crossing orbit in the past (\cite{mic98}).
Numerical integrations of Eros-type clones show that secular resonances
$\nu_4$ and $\nu_{16}$ probably modified Eros's orbital parameters, decreasing
its eccentricity enough to place it out of reach of the Earth, while
increasing Eros's inclination to its current value
(\cite{mic96,mic97a,mic98}).  If true, Eros would have been prone to low
velocity Earth encounters (and tidal disruption) in some past epoch.  

\section {Conclusions}

 Current evidence implies that km-sized asteroids and comets are rubble piles.
 When these objects, in the form of ECOs, encounter a planet like the Earth,
S-, B-, and M- class tidal disruptions frequently produce elongated objects
($\varepsilon_{{\rm rem}} > 0.6$) with fast spin rates ($P \sim 5$ hours). 
These values are consistent with at least two objects in near-Earth space,
1620 Geographos and 433 Eros, which may have made a close slow encounter with
Earth or Venus in the past. In addition, the shapes of our model asteroids
that have been heavily distorted (and disrupted) by Earth or Venus's tidal
forces resemble the radar-derived shapes of Geographos and Eros. Estimates of
the frequency of tidal disruption events indicate that a small but detectable
fraction of the ECO population should have Geographos-like spins and shapes. 
For these reasons, we believe that planetary tidal forces should be added to
collisional processes as recognized important geological process capable of
modifying small bodies.  

\acknowledgments

We thank L. Benner, J. A. Burns, P. Farinella, S. Hudson, J. Jenkins, S.
Ostro, and C. Thornton for useful discussions and critiques of this work.  We
also thank E. Asphaug and C. Chapman for his constructive reviews of this
manuscript.  PM worked on this paper while holding the External Fellowship of
the European Space Agency. DCR was supported by grants from the NASA
Innovative Research Program and NASA HPCC/ESS. Preparation of the paper was
partly supported by NASA Grant NAGW-310 to J. A. Burns.

\clearpage








%




%







%








\clearpage

\figcaption{1620 Geographos's pole-on shape determined from delay-Doppler
observations taken in the asteroid's equatorial plane (Ostro et al.,\ 1996). 
This image has been constructed from  multi-run sums of twelve co-registered
images, each $30^\circ$ wide in rotation phase space. The central white pixel
indicates the body's center-of-mass. Rotation direction is indicated by the
central circular arrow. Brightness indicates the strength of radar return,
arbitrarily scaled. Despite substantial smearing of the periphery features,
some distinguishing characteristics can be observed: (i) The long axis is
tapered on both ends, with one tip narrow and the other more pinched and
squat. (ii) One side is smooth and convex; the opposite side has a ``hump''.
(iii) Cusps at each end are swept back against the rotation direction, giving
the body the appearance of a pinwheel when viewed from various aspect angles.
The insets show close-ups from three of the twelve summed co-registered
$30^\circ$ images used to make the composite image; they have resolution of
500 ns $\times$ 1.64 Hz ($75 \times 87$ m). The cusps are more prominent here,
though considerable smearing remains.  \label{fig1}}

\figcaption {The ellipticity values of our model asteroids plotted against the
fraction of mass shed in each tidal disruption outcome. 79 S-class, 40
B-class, and 76 M-class disruptions are shown. The starting ellipticity for
our model rubble pile is $\varepsilon_{{\rm rem}} = 0.43$. Geographos's
ellipticity $\varepsilon_{\rm rem} = 0.64$. Note that most M-class disruptions
produce high ellipticities. \label{fig2}}

\figcaption{The final spin periods of our model asteroids plotted against the
fraction of mass shed in each of the 195 S-, B-, and M-class outcomes. 
Starting spin periods are $P$ = 4--12 h and $P= \infty$ (i.e.,  no spin). Note
that three S-class and one M-class events have final spin periods between
10--20 hours (i.e., beyond our $P = 10$ h plotting limit).  Regardless of the
starting spin period, most disruptions spin-up the model asteroid to $P < 6$
h. \label{fig3}}

\figcaption{Four snapshots of the tidal breakup by the Earth of a $P = 6$ h
prograde rotating rubble pile having $q = 2.1 R_\oplus$ and $v_\infty = 8$ km
s$^{-1}$. (a) shows the asteroid before encounter.  (b) shows the body shortly
after perigee passage. (c) shows the latter stages of tidal disruption as the
body recedes from the Earth. Particles shed near the tips do not return to the
rubble pile. (d) shows the final shape of the object.  Its spin ($P=5.03$ h)
and elongation ($\sim 2.9$ times the mean diameter of the minor axes, or
$\varepsilon_{{\rm rem}} \sim 0.65$) are virtually identical to Geographos
(Fig.\ 1). Spiral distortion associated with tides produces a smooth convex
surface along the long axis, cusps on either end, and a ``hump''-like mound of
material on the opposing side. \label{fig4}}

\figcaption[Eros.eps]{Pole-on silhouette of Eros, based on a model where radar
data were fit to a reference ellipsoid using 508 triangular facets defined by
256 vertices (Mitchell et al., 1998). The silhouette is viewed from the
asteroid's south pole. Definitions for center of figure, center of rotation,
and rotation direction are given in Fig.\ 1. The body is tapered along its
length, with a smooth convex side on the right and one or more concavities on
the left, making it look something like a kidney bean. Resolution does not
permit interpretation of the concavities on the left side (i.e., whether they
are craters, troughs, or bends in Eros's shape). \label{fig5}}

\clearpage
\plotone{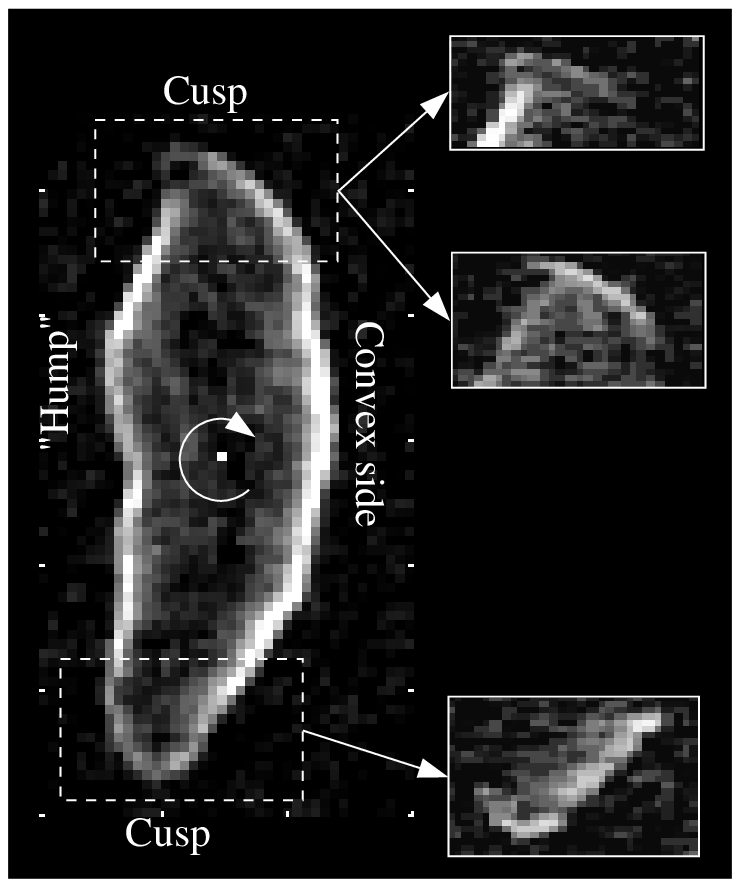}

\clearpage
\plotone{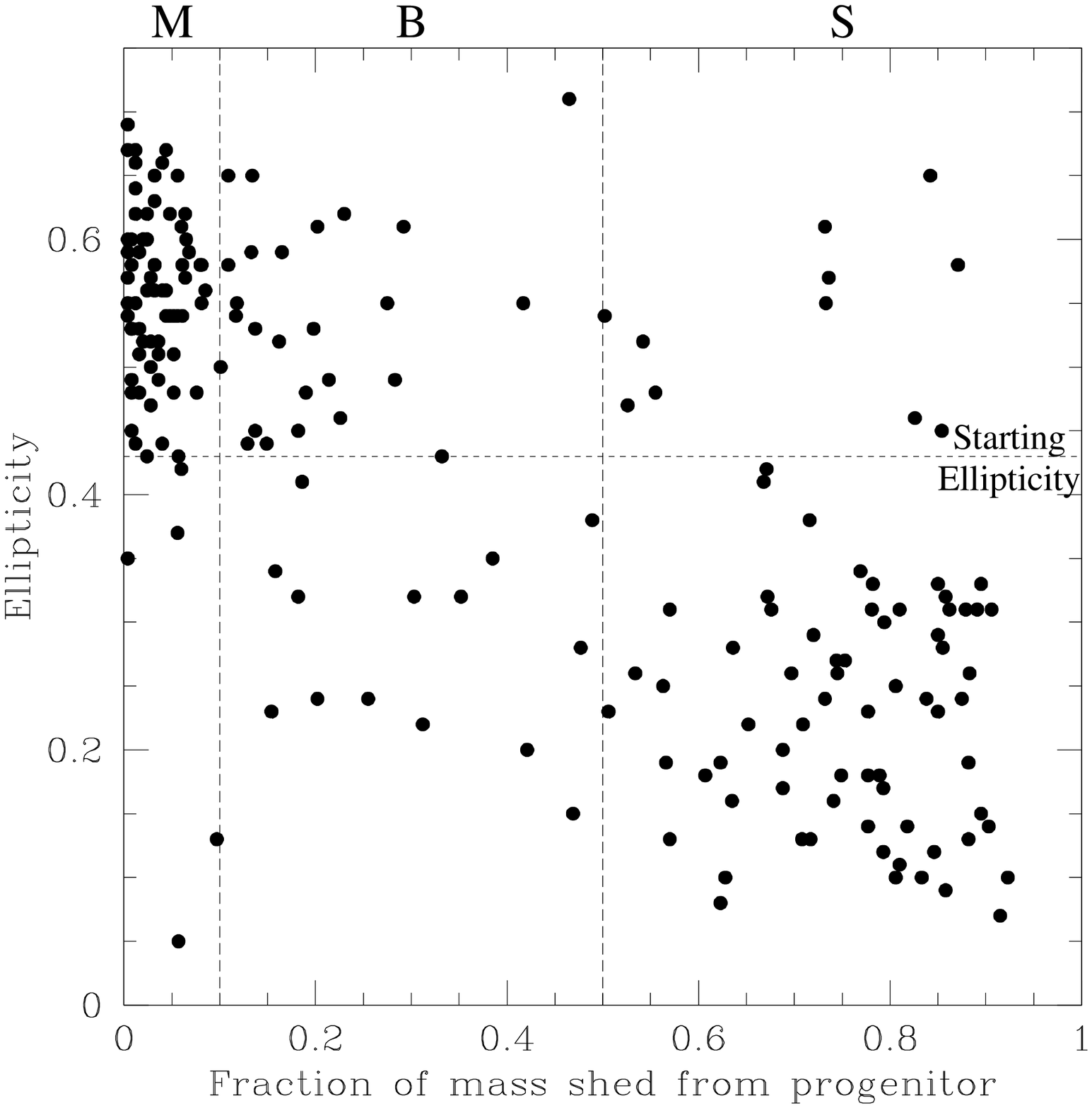}

\clearpage
\plotone{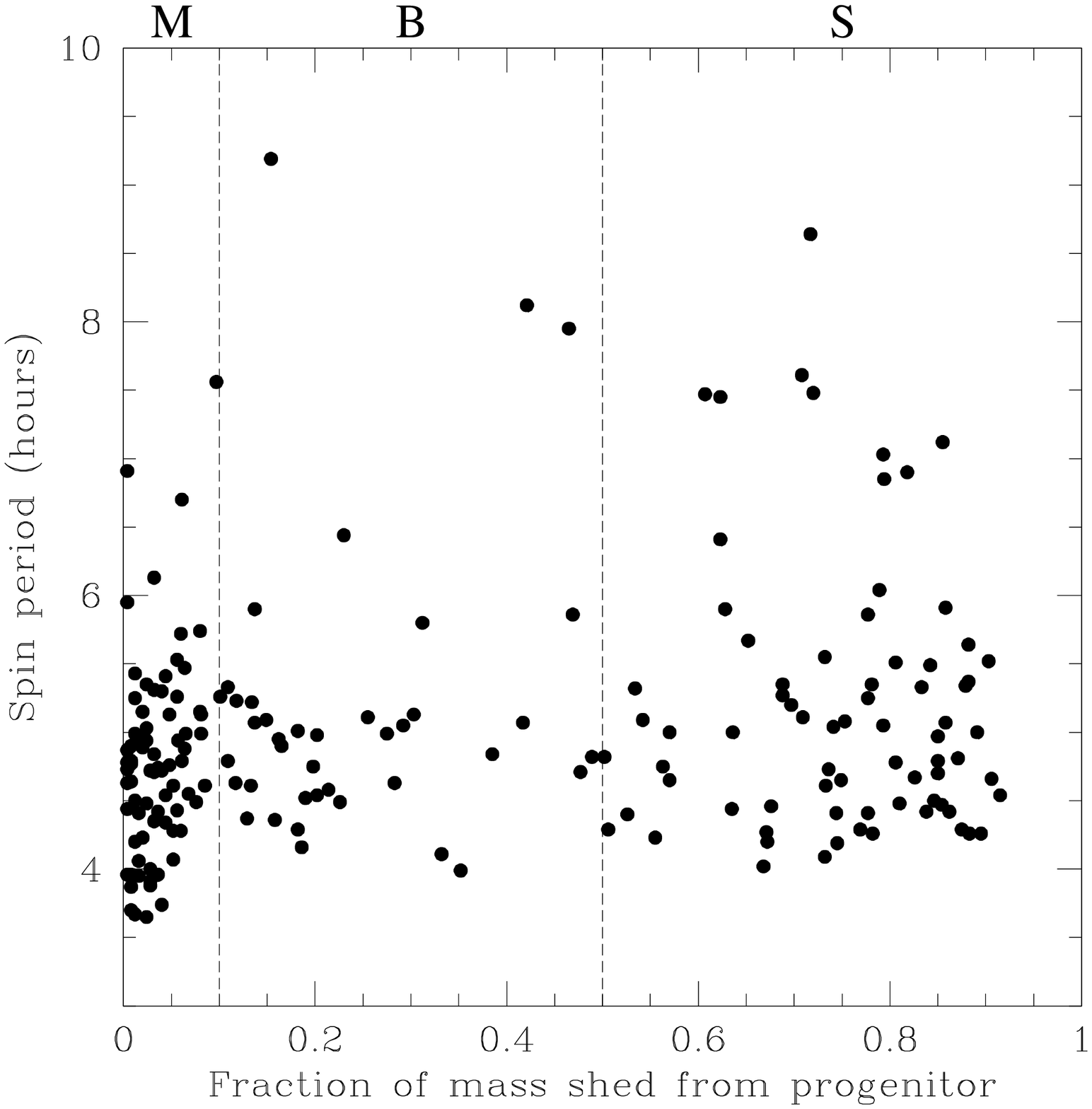}

\clearpage
\plotone{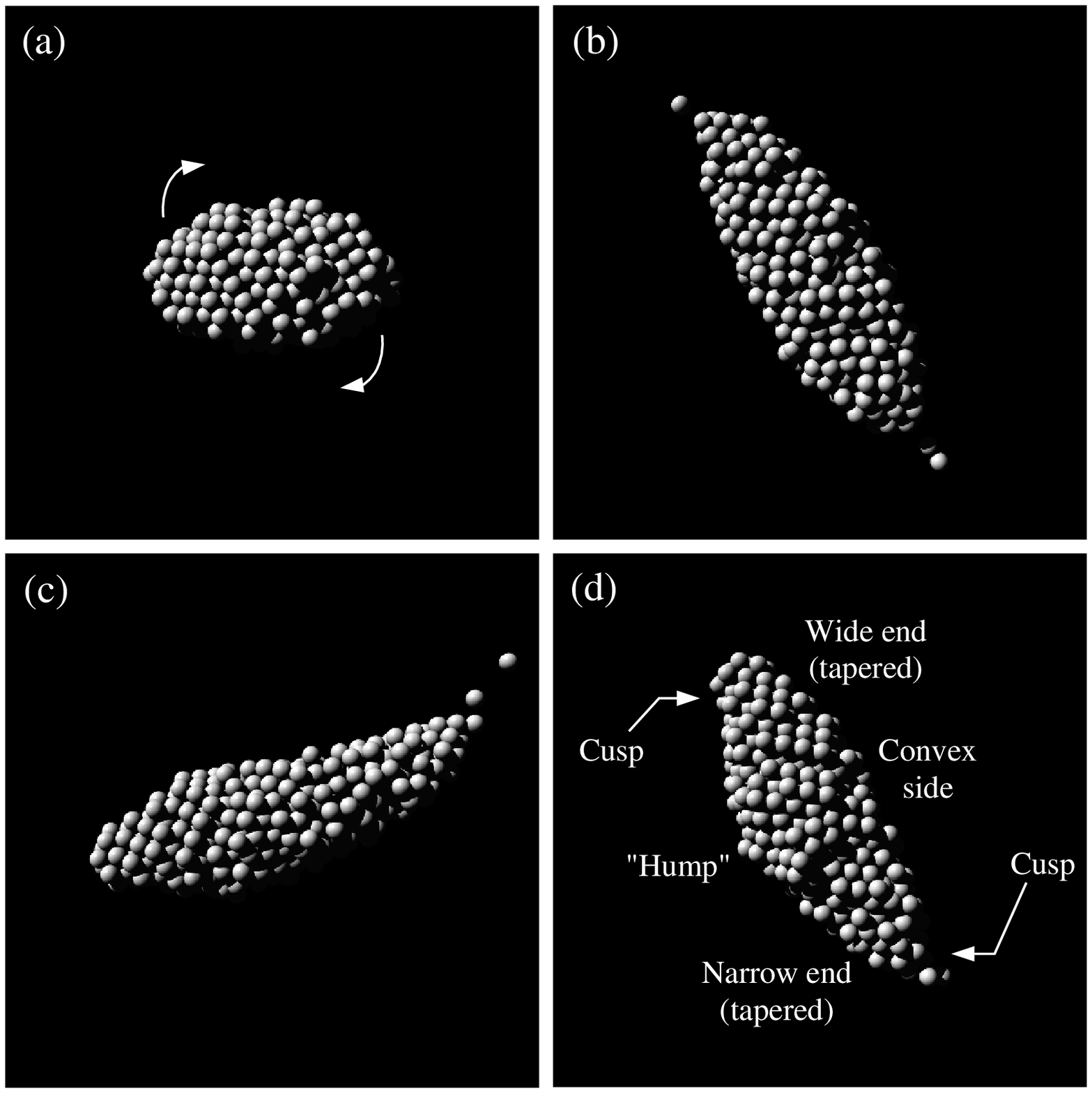}

\clearpage
\plotone{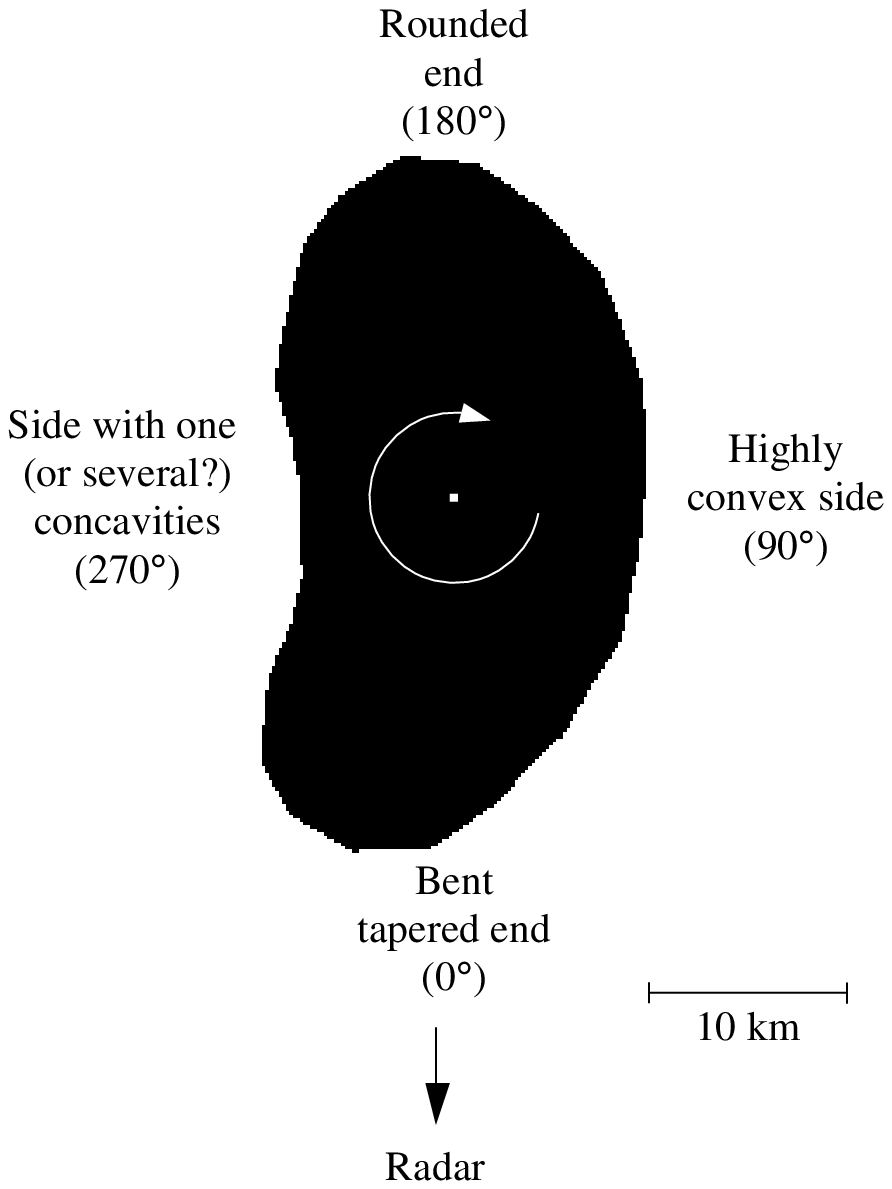}

\end{document}